# The genetic signature of (astronomically induced) life extinctions

[1]Robersy Sanchez and [2]Rolando Cardenas

[1] Centre for Studies on Bioinformatics, Universidad Central de Las Villas. Santa Clara. Cuba; and Research Institute of Tropical Roots, Tuber Crops and Banana (INIVIT), Biotechnology group, Villa Clara, Cuba.
E-mail: robersy@uclv.edu.cu

[2] Department of Physics, Universidad Central de Las Villas. Santa Clara. Cuba.
E-mail: rcardenas@uclv.edu.cu

**Abstract**

*The current understanding of supernova and gamma-ray burst events suggests important effects on the biosphere if one of more of them happened to strike the earth in the past. In this paper we evaluate the possibility that life extinctions which probably occurred due to excess of radiation occurring in the geologic past might have left a genetic signature on surviving species. We emphasize the signatures of these extinctions, proposing a quantitative model to evaluate the surviving probability of the species, based on kinetic aspects of the frequency of mutations and the DNA repair rate.*

## 1. Introduction

The most frequently invoked extraterrestrial mechanisms for life extinctions in the geologic past are bolide's impacts (e. g. comets from the Oort cloud) and nearby supernovae [1, 2]. Cosmic ray jets present in other astrophysical scenarios, like accretion discs and neutron star mergers, have been also proposed [3, 4], as well as gamma-ray bursts [2, 16] In all cases the aftermath implies a situation in which high levels of radiation are likely to reach the planet ground due to a handful of effects. Galante and Horvath [5] have classified the atmospheric/biospheric effects in:

a) Direct irradiation, given by the fraction of energetic photons that reach the ground;
b) UV retransmission, as part of the initial radiation is retransmitted by Compton scattering to the UV domain;
c) Ozone layer depletion, which allows a greater amount of solar UV radiation to reach the ground and;
d) Cosmic ray jets, which can produce muon showers even on deep underground environments through the atmospheric reaction:

$$p + N \rightarrow \pi + K \rightarrow \mu + \nu$$

The shock wave of impacts, or the gamma and cosmic ray bursts from supernovae, will considerably deplete the ozone layer, resulting in a high level of solar UV radiation in the ground. Cosmic ray bursts (CRB), through their interaction with the atmosphere, will additionally produce a muon shower at sea level, capable of a deep underwater and underground penetration (hundredths of meters).

The exact biological signature of extinctions may be quite complicated and somewhat obscured in fossil records. A first examination of the fossils suggest, however, a clear correlation between the extinction pattern of different species, their vulnerability to ionizing radiation and the sheltering provided by their habitats and the environment they live in. For instance, insects, which are less vulnerable to radiation (by a factor of 10 or 20 respect to vertebrates), were extinct only in the greatest extinction: the end Permian, 251 Myr ago. Even then only 8 out of 27 orders were extinct compared with a global extinction rate of around 95%. Mountain shadowing, canyons, caves, underground habitats, deep underwater habitats and high mobility may also explain why many species like crocodiles, turtles, frogs, snakes, deep sea organisms and birds were little affected in the Cretaceous/Tertiary boundary extinction which claimed the life of the big dinosaurs and pterosaurs [4].

Mutations continuously occur in the genomes of all living organisms, at a very low frequency, which tends to be constant for a specific species. That frequency is increased when the living organism is exposed to radiations. Highly efficient mechanisms of DNA repair explain the fact that in some areas with a relatively high level of radiation, harmful effects on the biota are not noticeable [6]. For instance, recently was identified a gen (*irrE*), which contributes to a exceptionally high resistance of the bacteria *Deinococcus radiodurans* to radiation [7]. This organism beats most of the constraints for survival of life on Mars - radiation, cold, vacuum, dormancy, oxidative damage, and other factors [8]. Evidences suggest that DNA repair mechanisms have been preserved during evolution, so it is very probable that genes like *irrE* might also be found in humans.

In this paper we present additional considerations on the surviving/death probability of the species exposed at high levels of radiation, considering the DNA repair mechanisms. Considering the fact that by 3500 million years ago, most of the principal biochemical pathways that sustain the modern biosphere had evolved and were global in scope [9], we think that our model can shed some more light on surviving chances on species exposed to high levels of radiation in the geologic past.

## 2. The Model

All cells have repair systems to protect themselves against damage from the environment or errors that may occur during replication. The overall mutation rate is the result of the balance between the introduction of mutations and their removal by these systems. Our model is based on three simple kinetic postulates:

I) The mutation rate due to radiations ($V_{rad}$) is directly proportional to the applied radiation dosage ($E_{rad}$):

$$V_{\mathrm{rad}} = k\, E_{rad} \tag{1}$$

II) The rate of enzymatic DNA repair mechanisms is determined by the rate of the slowest enzymatic reaction.

III) In the DNA molecule, in environmental conditions allowing the viability of living organisms, the change in time of the frequency of mutations ($N(t)$) is equal to zero:

$$\frac{dN(t)}{dt} = 0 \tag{2}$$

The first two postulates are very simple and quite intuitive. The third postulate means the existence of a stationary state in stable environmental conditions for life; i.e. in this environment the frequency of mutations in the DNA molecule $N(t)$ is constant, $N(t) = N$. The mutation frequency $N(t)$ can be given as a linear concentration, i.e. the number of mutations per base pair in the DNA molecule or the number of mutations per locus.

Following a classical enzymatic analysis we assume that DNA repair reactions satisfy the Michaelis-Menten equation

$$V_{rep} = \frac{V_{\max} N(t)}{K_M + N(t)} \quad (3)$$

where, $V_{rep}$ is the DNA repair rate, $V_{max}$ is the maximum rate of the enzymatic reaction, $N(t)$ is the linear concentration of mutations in the DNA molecule and, $K_M$ is the Michaelis constant.

Now, assuming that radiations are the main cause of mutations, the approximation of stationary state gives the equation:

$$V_{rad} - V_{rep} = \frac{dN(t)}{dt} = 0 \quad \mathbf{(4)}$$

Combining expressions (1), (3) and (4) we have:

$$N(t) = N = \frac{k\, E_{rad}\, K_M}{V_{\max} - k\, E_{rad}} \quad \mathbf{(5)}$$

Obviously, if the mutation rate due to radiations is greater than the maximum repair rate ($V_{max} < kE_{rad}$), the equation (5) looses its validity. Actually, organisms die due to the excess of non repaired mutations. However, when the maximum rate of the enzymatic reactions $V_{max}$ is much greater that the mutation rate due to radiations ($V_{max} >> kE_{\mathrm{rad}}$), we have:

$$N = \frac{k\, K_M}{V_{\max}} E_{rad} = c\, E_{rad} \quad \mathbf{(6)}$$

That is, the frequency of mutations grows linearly with the radiation dosage in some interval $\{0, V_o\} : 0 < kE_{\mathrm{rad}} < V_o < V_{max}$.

## 3. Experimental Confrontations

The equation (6) is essentially confirmed by the results of Russell and Kelly [10] who obtained the mutation frequencies in male mice. If we take the control of the mutation frequencies as baseline, then the regression analysis lead to a satisfactory fit for this linear approximation (see Fig 1).

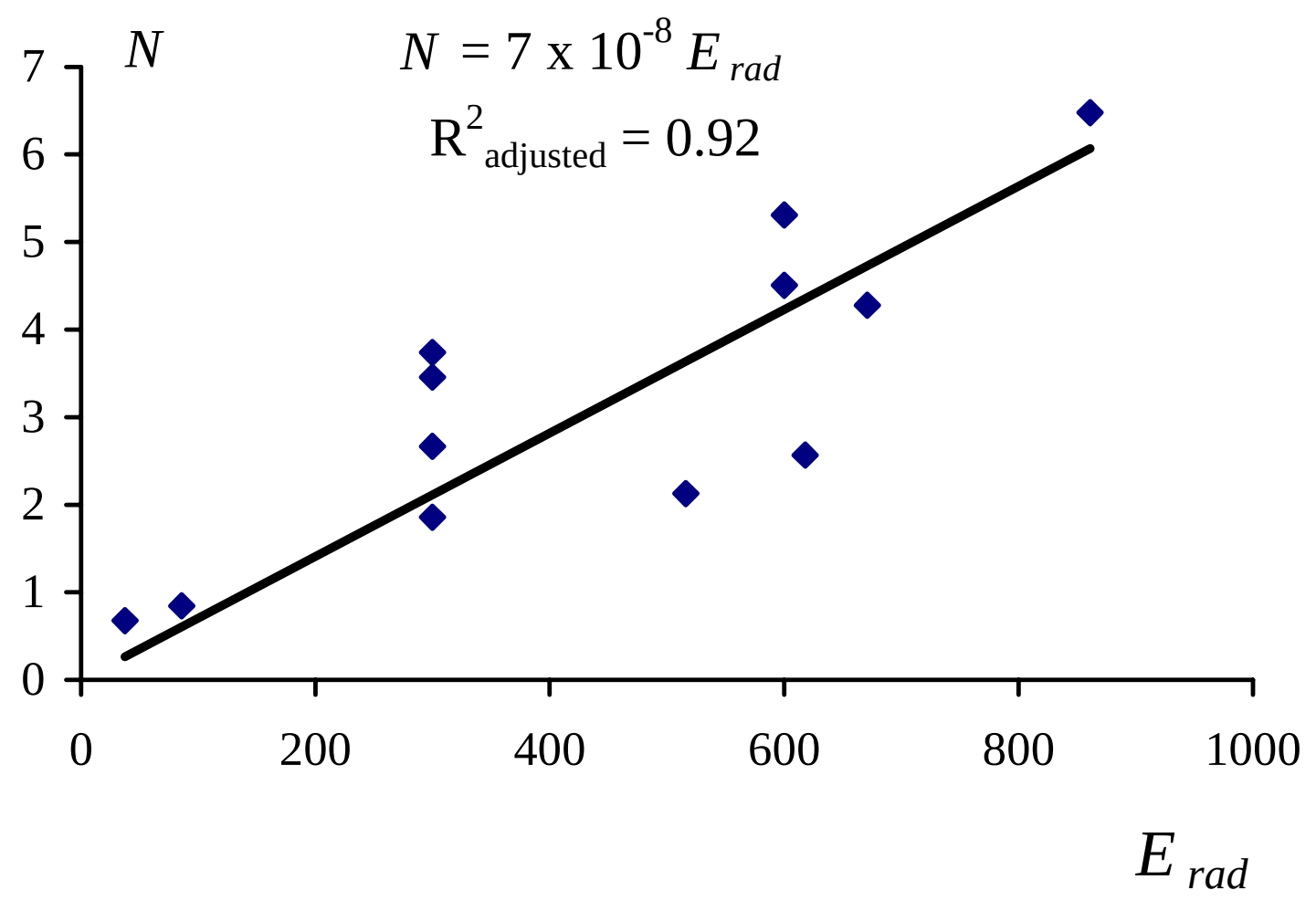

**Figure 1.** Graph of the regression analysis of the frequency of mutations *N* versus dose ($E_{rad}$). The $f_r$ values correspond to the number of mutation × $10^{-5}$ per locus and the values $E_{rad}$ are in Roentgens. All data were taken from Table 1 of [10]. The values *N* were readjusted taken the mean of control values as baseline in the regression through the origin. The 95% confidence interval for the regression coefficient is: lower bound equal to 5.69 × $10^{-8}$ and upper bound equal to 8.39 × $10^{-8}$.

### 3.1. The estimation of surviving chances

Survival depends upon the ability of organisms to cope with varied environmental stress conditions, including fluctuating water availability, heat, osmotic stress, and radiation. Here, we only have taken into account radiation stress.

Suppose the simplest relationship between the survival *S* of irradiated organisms and the frequency of mutations *N*. That is, *S* is assumed inversely proportional to *N*; this means that:

$$S = \frac{s}{N} = \frac{sV_{\max}}{K_M k}\frac{1}{E_{rad}} - \frac{s}{K_M} \qquad \textbf{(7)}$$

We obtain again the equation of a hyperbola, or that of a straight line if we plot *S* against the inverse of the radiation dosage $1/E_{rad}$.

#### 3.1.1 Effect of UV radiation

A report [11] about the effect UV radiation on the bacteria of the plant leaf surface, termed the phyllosphere, suggests the validity of the equation (7). Phyllosphere microbial residents grow through the utilization of the limited resources available in this habitat. Jacobs and Sundin [11] pointed out that solar UV-B selection alters phyllosphere bacterial community composition and that UV tolerance is a prevalent phenotype late in the season. They selected *Clavibacter michiganensis* as a model UV-tolerant epiphyte in their study. Here, we plot the survival of *Clavibacter michiganensis* on peanut leaves under the effect of solar UV-B radiation versus the inverse UV of radiation dosage $1/E_{rad}$. In Figure 2 the linear regression analyses of the survival percent at different times are shown: 24, 48, 72 and 96 hours after inoculation. The data were taken from Table 5 of [11]. Note that we can estimate the vertical asymptote $V_{max}/k$ of equation (5) using equation (7). This magnitude is a measure of repair capacity of the cells and consequently a measure of its radiation resistance, therefore a large value of $V_{max}/k$ assures a high survival.

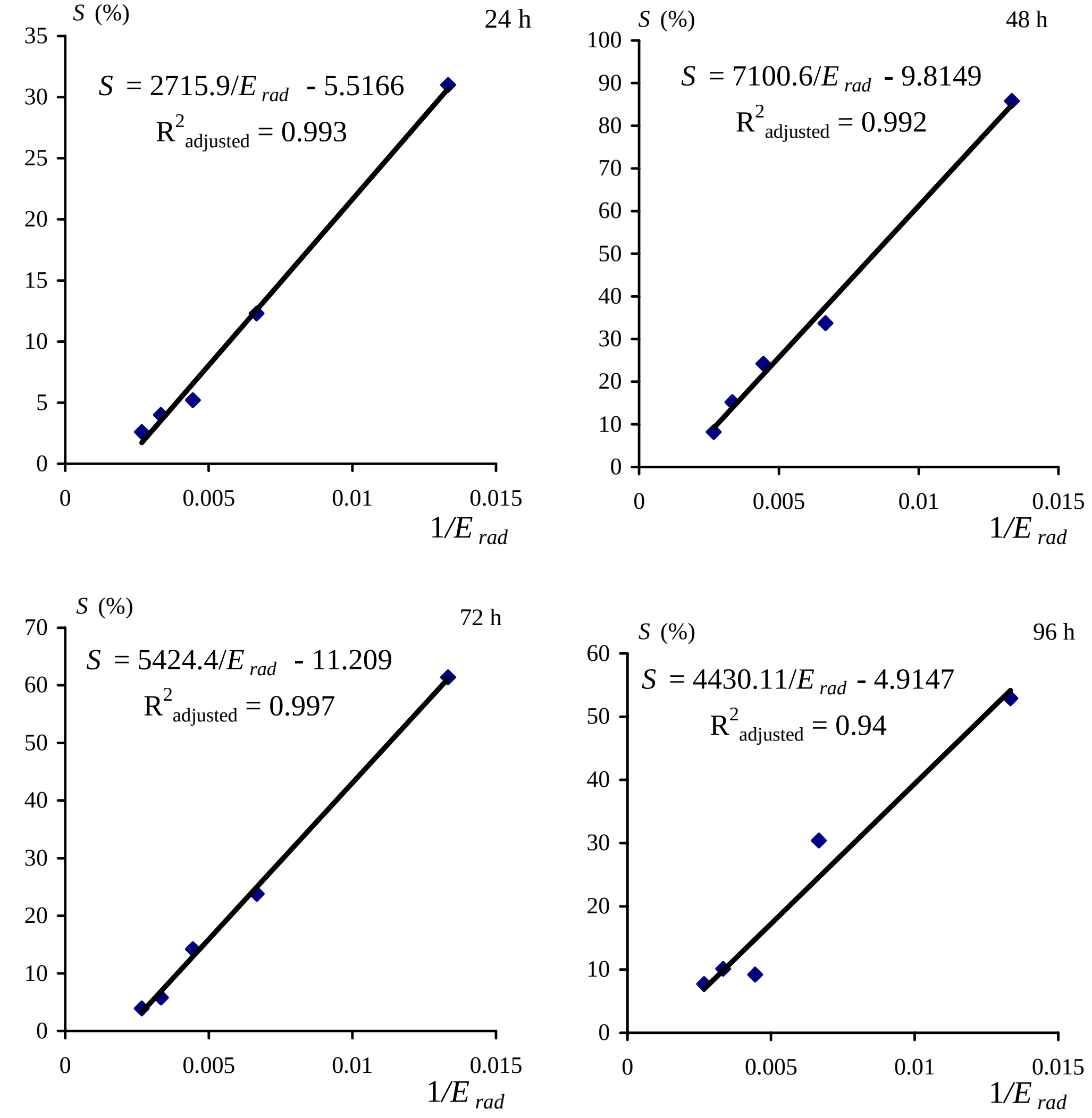

**Figure 2**. Graphs of the linear regression analyses of the survival of *Clavibacter michiganensis* versus the inverse UV of radiation dosage $1/E_{rad}$. The regression analyses of the survival percent were made for values at different times: 24, 48, 72 and 96 hours after inoculation. The data were taken from Table 5 of [11]. The values of $V_{max}/k$ at different times are: 492.31 (24 hours); 723.45 (48 hours); 483.93 (72 hours) and 901.40 (96 hours).

In present case the values of $V_{max}/k$ at different times are: 492.31 (at 24 hours); 723.45 (at 48 hours); 483.93 (at 72 hours) and 901.40 (at 96 hours). This means that radiation effects on cell vary the enzymes composition of the enzymatic reparation systems after initial radiation moment, and the maximal efficiency is not reached in the first hours.

### 3.1.2 Effect of ionizing radiations

A behavior qualitatively similar to above is observed when gamma radiation ($^{60}Co$) is employed as the irradiation source. In Fig 3 is showed the regression analysis of the response of freshly hatched larvae of *Trogoderma Granarium* everts (% Survival) versus the inverse of gamma radiation dose ($1/E_{rad}$). The values were taken from [12].

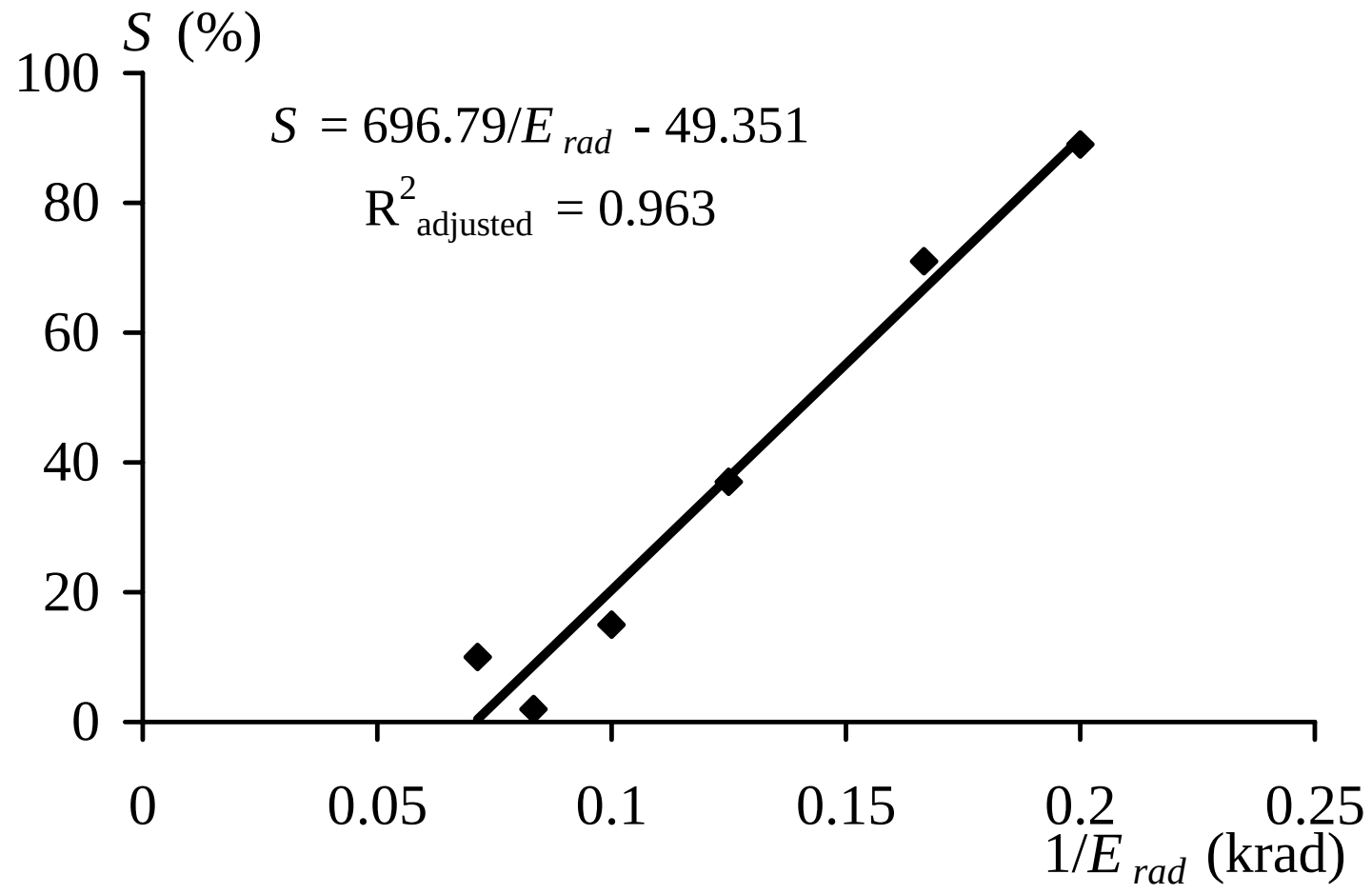

**Figure 3**. Regression analysis of the response of freshly hatched larvae of *Trogoderma Granarium* everts (% survival $S$) versus the inverse of gamma radiation dose ($1/E_{rad}$). Data were taken from [12].

Also, for the survival of Carica Papaya irradiated with gamma rays ($^{60}Co$) we get a satisfactory fit. In the Fig 4 it is shown the regression analysis of the percent of explants survival versus inverse of dose (the data are taken from the genetic improvement program at Research Institute of Tropical Roots, Tuber Crops and Banana (INIVIT), Villa Clara, Cuba , J. Ventura personal communication).

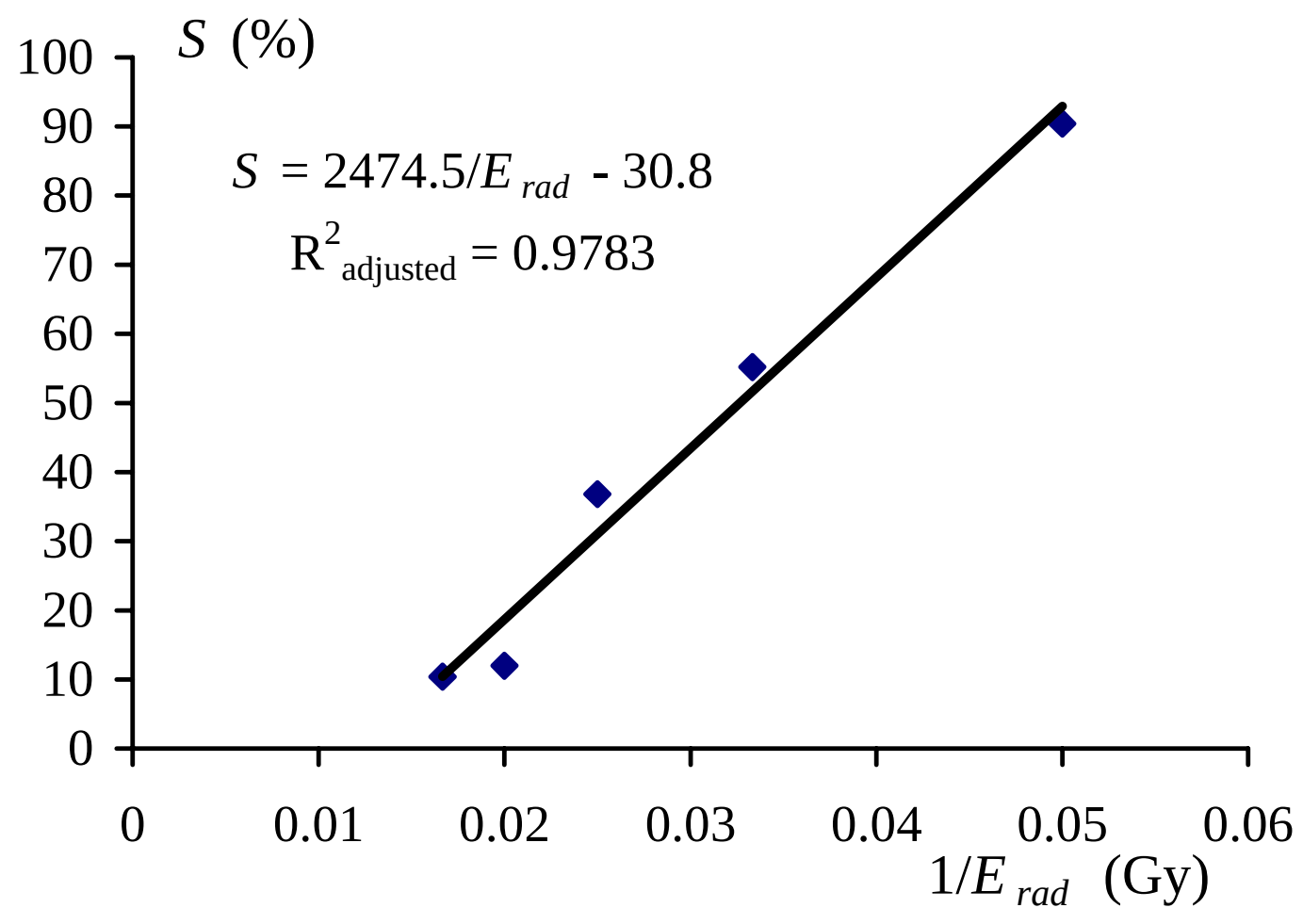

**Figure 4.** Regression analysis percent of the *Carica papaya* explants survival $S$ versus inverse of dose $1/E_{rad}$.

A behavior qualitatively similar to above seems to be obtained for the case of irradiation with X-rays, in our work in progress.

## 4. Discussion

From experimental confrontations of equations (5) and (7), it is deduced that the rate $V_{max}/k$ may be used as a measure of the repair capacity of the cells and consequently a measure of its radiation resistance. In particular, we see that the rate $V_{max}/k$ is a kinetic constant characteristic for every enzymatic system. On the cell the radiation-induced variation of the rate $V_{max}/k$ corresponds to the variation in the enzymatic composition of the repair systems after radiation. Actually, this is an expected result: repair genes can be classified into pathways that use different mechanisms to reverse or bypass damage to DNA. The enzymes associated with these systems correspond to several repair pathways [13].

Obviously, if $E_{rad}$ increases slowly near to the current $V_{max}$ , it should be possible to pass from one stationary state to another. A gradually increase or small jumps of the radiation dose should induce new enzymes from different repair pathways. Perhaps this observation could help to explain why mutation rate tends to keep constant in those populations with high background radiation [6]. All cells have many pathways to repair damage in DNA, which pathway is used will depend upon the type of damage and the situation [13].

The mutation rate reflects a balance between the number of damaging events occurring in DNA and the number that have been corrected. Given any astrophysical situation in which high levels of radiation reach the ground, mutations will be accumulated instead of being removed in those species for which the mutation rate due to radiations is greater than the maximum repair rate ($V_{max}< kE_{rad}$). As a result, if these species have evolved in an exposed environment, they will be extinct.

Of course, it is well know that repair systems are as complex as the replication apparatus itself, which indicates their importance for the survival of the cell. Therefore, our model is a kinetic approximation to describe mutation damage in DNA. Actually, some deviation of our model is to be expected at early times [14], which can be explained considering that it takes some time to reach the stationary state described by equation (4). It should also be noted that the model is not to be applied to low doses situations, in which survivence is practically not affected.

## 5. Conclusions

As can be seen from above examples involving mice, phyllosphere and Carica Papaya, the proposed kinetic model reproduces satisfactorily, from a genetic point of view, and for a diversity of species, the surviving chances when living organisms are exposed to radiations, both UV and gamma.

Radiations can enhance the molecular evolution when the mutation rate is not greater than the maximum DNA repair. Such a situation might have been the case for Cambrian and Late Ordovician mass extinction, which, according to some authors might have been caused by gamma-ray bursts [2, 15]. In these cases, the high level of radiation in the biosphere would have played the double role of sterilizing those species for which the mutation rate exceeds the DNA repair rate, and allowing surviving those in which repair rate would be greater than mutation rate, enhancing the speciation due to mutations and empty ecological niches. We suggest that such a behavior might open a road to understand the so-called Cambrian explosion [2].

## Acknowledgements

We acknowledge CAPES from Brazil and MES of Cuba for financial support. We also acknowledge Jorge Horvath, from University of Sao Paulo, for helpful comments and discussions.